\documentclass[letter]{aa}
\usepackage[sort]{natbib}
\bibpunct{(}{)}{;}{a}{}{,}  % to follow the A&A style
\usepackage{graphicx}
\usepackage{txfonts}

\def\Ks{{K$_\mathrm{s}$}}

\def\Brg{{Br$_\gamma$}}
\def\MK{{M$_\mathrm{K}$}}
\def\micron{$\mu$m}
\def\kms{{km sec$^\mathrm{-1}$}}

\begin{document}
\title{Pattern speed of main spiral arms in NGC~2997\thanks{Based
    on observations made with ESO telescopes at the La Silla Paranal
    Observatory under programme ID 278.B-5042}}
\subtitle{Estimate based on very young stellar complexes}
\titlerunning{Estimation of pattern speed in NGC~2997}
\author{P.~Grosb{\o}l\inst{1} \and H.~Dottori\inst{2}}
\offprints{P.~Grosb{\o}l}
\institute{European Southern Observatory,
  Karl-Schwarzschild-Str.~2, DE-85748 Garching, Germany\\
  \email{pgrosbol@eso.org}
\and
  Instituto de F\'{i}sica, Univ. Federal do Rio Grande do Sul,
  Av. Bento Gon\c{c}alves~9500, 91501-970 Porto Alegre, RS, Brazil\\
  \email{dottori@ufrgs.br}
}

\date{Received ??? / Accepted ???}

%-----------------------------------------------------------------
\abstract{Bright knots along arms are observed in K-band images of many
  grand-design spiral galaxies. They are identified as young starforming
  complexes using K-band spectra which show strong {\Brg} emission.
  Their alignment with spiral arms and ages $<$10~Myr suggest that they are
  associated with a starforming front linked to a density wave in the
  galaxies. }
{Ages may be estimated for the youngest starforming complexes using NIR broad
  band colors and \Brg.  A different angular speed of the density wave
  and material would lead to an azimuthal age gradient of newly formed
  objects. We aim to detect this gradient. }
{Deep JHK-{\Brg} photometry of the southern arm of the grand-design
  spiral galaxy NGC~2997 was obtained by ISAAC/VLT. All sources in the field
  brighter than K=19~mag were located.  Color-color diagrams were
  used to identify young stellar complexes among the extended sources. Ages
  can be estimated for the youngest complexes and correlated with azimuthal
  distances from the spiral arms defined by the K-band intensity variation.  }
{The extended sources with \MK $<-12$~mag display a diffuse appearance and
  are more concentrated inside the arm region than fainter ones, which are
  compact and uniformly distributed in the disk.  The NIR colors of the bright
  diffuse objects are consistent with them being young starforming complexes
  with ages $<$10~Myr and reddened by up to 8 mag of visual extinction. They
  show a color gradient as a function of their azimuthal distance from the
  spiral arms.  Interpreting this gradient as an age variation, the pattern
  speed $\Omega_\mathrm{p}$ = 16 km sec$^{-1}$ kpc$^{-1}$ of the main spiral
  was derived assuming circular motion. }
{The alignment and color gradient of the bright, diffuse complexes strongly
  support a density wave scenario for NGC~2997.  Only the brightest complexes
  with \MK\ $<-12$~mag show a well aligned structure along the arm, suggesting
  that a strong compression in the gas due to the spiral potential is required
  to form these most massive aggregates, while smaller starforming regions are
  formed more randomly in the disk. The sharp transition between the two
  groups at \MK\ $= -12$~mag may be associated with expulsion of gas when the
  first supernovae explode in the complex. }
\maketitle
%-----------------------------------------------------------------
\section{Introduction}
\citet{stromgren63} suggested the use of stellar ages and velocities to
estimate properties of the spiral potential in our Galaxy, assuming that stars
are formed preferentially in spiral arms.  Based on this assumption,
birthplaces of B-stars in the solar neighborhood with ages in the range of
100--300~Myr were calculated \citep{stromgren67, yuan69, grosbol76} to
determine pattern speed and amplitude of a possible density wave
\citep{lin64}.

For other galaxies, one may consider integrated quantities since neither
accurate ages nor space velocities can be observed for individual stars.  The
standard density wave picture suggests that stars are more likely to be formed
in spiral arms \citep{roberts69} that rotate with a constant angular speed
$\Omega_\mathrm{p}$.  This would lead to an age gradient in young objects as
the density wave moves relative to the material \citep{yuan81}. Broad band
color gradients across spiral arms have been reported \citep{gp98, puerari97,
  gonzalez96} although the interpretation may be complex due to significant
attenuation by dust in the arm regions.  Phase shifts between spiral arms
defined by H$\alpha$ and CO emission were found by \citet{egusa04} who
attributed this to differential motion between the density wave and
interstellar material.

Deep images of spiral galaxies in the near-infrared (NIR) K-band \citep{gp98}
revealed bright knots along the arms of several grand-design spirals.  Their
compactness and alignment suggested that they are young objects.  Low
resolution K-band spectra of such knots in NGC 2997 \citep{grosbol06} showed
{\Brg} emission and clearly identified them as very young stellar complexes
with ages $<$10~Myr.  A correlation between {\Brg} emission (i.e. age) and
azimuthal distance from the spiral arm of the 6 individual complexes
investigated suggested that their formation was triggered by a front
associated with a density wave.

Synthetic spectra of very young clusters from {\sl starburst99} \citep[
  hereafter SB99]{leitherer99} suggest that one can estimate their ages from
NIR broad-band colors even with significant reddening.  In this paper, we
present deep JHK-{\Brg} photometry to determine if an age gradient can be
detected for the brightest knots in NGC~2997 and, based on that, derive an
estimate of the pattern speed of the underlying density wave.

\section{Data and reductions}
\begin{figure}
  \resizebox{\hsize}{!}{\includegraphics{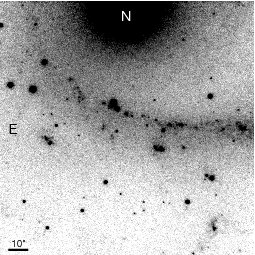}}
  \caption[]{Direct K-band image of NGC~2997 in intensity scale where the bar
  indicates scale.}
  \label{n2997k}
\end{figure}

The spiral galaxy NGC~2997, with Hubble type Sc(s)I.3, is one of the
nearest grand-design spirals that displays a strong, symmetrical two-armed
pattern with a large number of bright complexes aligned along the arms
\citep{grosbol08}.  A field including its southern arm (see Fig.~\ref{n2997k})
was observed with ISAAC, which has a Rockwell Hawaii 1024$\times$1024 Hg:Cd:Te
detector and a pixel scale of 0.148\arcsec.  The service mode observations
were done at VLT in the period Feb. 9-11, 2007, with exposure times of
4$^\mathrm{m}$, 4$^\mathrm{m}$, 10$^\mathrm{m}$, and 25$^\mathrm{m}$ on-target
in the J, H, \Ks, and {\Brg} filters, respectively.  The same observing
template was used for all filters and specified a jitter pattern around the
target position with offsets around 10\arcsec\ interleaved with sky exposures
significantly outside the galaxy, avoiding bright stars.  The reductions were
done by an ESO-MIDAS based pipeline which used the standard calibration data
provided by the ISAAC calibration plan.  The seeing of the final, stacked
frames was 0.7\arcsec, corresponding to 40~pc adopting a distance modulus of
30.38 mag.

All discrete sources on the K frame were identified using {\sl SExtractor}
v2.5.1 \citep{bertin96} with a detection threshold of 2.5.  The mesh size was
reduced to 16 and local background estimates were specified in order to follow
the intensity variation across the spiral arm.  With these positions, aperture
photometry was performed on the 4 stacked frames using a 2\arcsec\ diameter
and the background computed in a ring with 3\arcsec\ radius. Individual errors
were estimated considering scatter in the backgound.  Magnitudes J=20.3,
H=19.6, K=19.3, and \Brg=19.0 were reached with errors $<$0.1 mag.

% J=20.3, H=19.6, K=19.3, and \Brg=19.0

\begin{figure}
  \resizebox{\hsize}{!}{\includegraphics{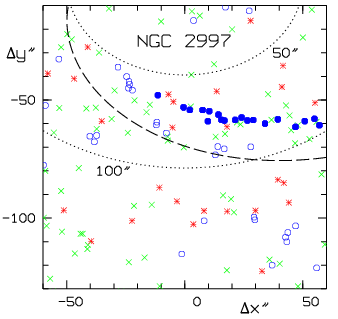}}
  \caption[]{Location of SExtractor sources identified on the K-band map of
    NGC~2997. Coordinates are given relative to the center of the galaxy.  The
    dashed line shows the location of the spiral arm. Symbols indicate the
    class of the objects: point-like ($\ast$), compact ($\times$), and diffuse
    ($\circ$) with filled circles indicating clusters in the main string along
    the southern arm.  The dotted ellipses outline circles in the plane of the
    disk with radii of 50 and 100~pc using a sky projection (PA,IA) =
    (102\degr,38\degr).}
  \label{n2997xy}
\end{figure}

Sources were grouped into 3 classes using the {\sl class\_star} index (cs) of
{\sl SExtractor}, which ranges from 1.0 for point-like sources to 0.0 for
extended ones.  Targets with cs$>$0.8 were assumed to be foreground stars
whereas extended objects were separated into compact and diffuse sources
depending on their index being above or below 0.3.  The distribution of the
different classes relative to the spiral pattern is shown in
Fig.~\ref{n2997xy}. The diffuse sources show a strong concentration in the
spiral arm with a long, well aligned string of complexes (filled circles)
15-25\arcsec\ inside the K-band arm defined by its m=2 Fourier component.
Both foreground stars and compact objects have a more uniform spatial
distribution.

\section{\bf Colors and ages of extended sources}
\begin{figure}
  \resizebox{\hsize}{!}{\includegraphics{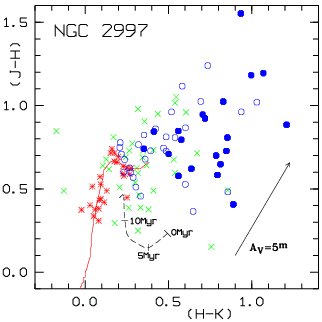}}
  \caption[]{Color-color diagram of objects in the NGC~2997 field using the
    infra-red colors (H-K) -- (J-H). Only extended sources with an error less
    than 0.1 mag in their color index are plotted. Symbols are the same as in
    Fig.~\ref{n2997xy}. The full line indicates the location of the
    local stellar main sequence.  The dashed line shows the evolution of a
    stellar cluster using the continuous SB99 model.  The reddening
    vector in the lower right indicates the offset due to 5~mag of visual
    extinction.}
  \label{n2997-ccd}
\end{figure}

An indication of the age of the individual sources can be obtained from the
(H-K) -- (J-H) color-color diagram given in Fig.~\ref{n2997-ccd}, where all
objects with errors $<$0.1 mag in their color indices are shown.  The
point-like sources cluster around the stellar main sequence, suggesting that
the zero-point error in the color indices is $<$0.01 mag.  The analysis of NIR
spectra of complexes in NGC~2997 \citep{grosbol06} and galactic starforming
regions \citep{lee07} indicate that the age of the starforming regions is
better described by a continuous star formation scenario.  The evolutionary
track of a stellar cluster with continuous star formation is display in
Fig.~\ref{n2997-ccd} using an SB99 model with a Kroupa IMF, upper mass limit
of 120~M$_\mathrm{\sun}$, solar metallicity, and Padova isochrones with AGB
stars.  The track starts at (H-K, J-H) = (0.50, 0.23) and reaches (0.24, 0.31)
for an age of 10~Myr, after which color changes are slow, with old stellar
clusters around (0.2, 0.6).  Almost all extended sources scatter above the
track in the direction indicated by the galactic reddening vector in
Fig.~\ref{n2997-ccd}.  Thus, the majority of the diffuse knots are likely to
be stellar complexes or clusters with up to 8 mag of visual extinction.  This
is consistent with estimates based on \Brg\ and Pa$_\beta$ lines
\citep{grosbol06} but is significantly more than measures based on visual
lines \citep{walsh89}.  A detailed comparison with H$_\alpha$ maps, available
from VLT/FORS1, reveals that H$_\alpha$ emission is detected from many of the
K-band knots but has different morphology and therefore does not originate
from the exact same regions.

One may construct a first order 'reddening corrected' index Q = (H-K) -
0.59$\times$(J-H) using a screen model with standard galactic extinction
\citep{winkler97}.  SB99 models with continuous star formation show that age
correlates well with both Q and {\Brg} emission for clusters younger than
8~Myr (i.e. 0.1$<$Q). To verify this, a {\Brg} index $\Gamma$ = (\Brg-K) was
computed with a zero point defined by the average of the foreground stars.
Although a general correlation is seen in Fig.~\ref{n2997-cmd}, the Q index
was preferred as an age indicator due to the higher error in $\Gamma$. An
additional issue for the ISAAC {\Brg} filter is its steep red transmission
edge around 2.174\micron\ which is close to the {\Brg} line for the
heliocentric velocity of 1088 \kms\ for NGC~2997.

The distribution of absolute magnitude {\MK} as a function of Q shows a clear
separation of young sources with 0.1$<$Q where all diffuse ones are brighter
than -12 mag, while compact ones are fainter (see Fig.~\ref{n2997-cmd}).  Most
extended sources lie in the predicted range of -0.2$<$Q$<$0.4 for clusters
with ages $<$300~Myr. The few outliers may be caused by the actual reddening
law differing from the galactic one or the more complex geometry of stars and
gas in the complexes \citep{witt92, pierini04}, as found by \citet{grosbol08}
for NGC~5247 by comparing (H-K)--\MK\ and Q--\MK\ diagrams.

\section{Pattern speed of spirals}
The alignment of the diffuse sources along the spiral arm makes it interesting
to estimate the azimuthal phase distance between the individual knots and the
spiral pattern defined by the intensity variation in the K-band image as given
by \citet{grosbol04}.  The phase of the m=2 Fourier component in the plane of
the galaxy fits a logarithmic spiral well with a pitch angle i = 22.6$^\circ$
in the radial range of 45\arcsec$<$r$<$125\arcsec\ while i = 21.8$^\circ$ is
estimated for 65\arcsec$<$r$<$105\arcsec. 
The string of bright, diffuse complexes inside the main arm
fits a slightly more open spiral with i = 29$^\circ$.

\begin{figure}
  \resizebox{\hsize}{!}{\includegraphics{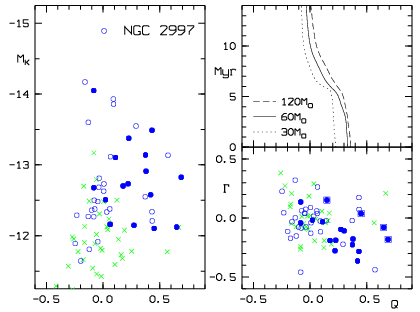}}
  \caption[]{Color-magnitude diagrams of non-stellar objects with errors
    $<$0.1 mag in the field of NGC~2997. {\sl The left plot} shows the
    absolute K magnitudes, \MK, as a function of the `reddening corrected'
    index Q.  {\sl The lower right} displays the {\Brg} index $\Gamma$
    as a function of Q where only sources are shown. {\sl The upper right}
     shows the Q--age relation for three SB99 models with upper mass
    limits of 30, 60, and 120~M$_\mathrm{\sun}$. Same symbols as in
    Fig.~\ref{n2997xy} are used; boxes indicate 4 diffuse clusters
    that do not follow the Q-$\Gamma$ relation. }
  \label{n2997-cmd}
\end{figure}

\begin{figure}
  \resizebox{\hsize}{!}{\includegraphics{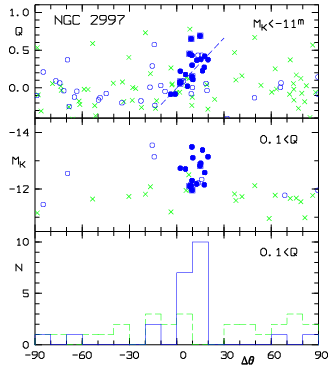}}
  \caption[]{Distribution of extended sources as a function of their azimuthal
    distance from the K-band spiral pattern. {\sl The upper diagram} shows the
    `reddening' corrected color index Q.  {\sl The middle one} displays the
    absolute magnitude M$_\mathrm{k}$ for young complexes with 0.1$<$Q; {\sl
      the lower diagram} shows the histogram. Same symbols as in
    Fig.~\ref{n2997-cmd}. }
  \label{n2997-dth}
\end{figure}

The azimuthal phase difference $\Delta\theta$ between the individual complexes
and the average logarithmic K-band spiral is shown in Fig.~\ref{n2997-dth},
where positive values indicate locations inside the spiral arm.  The histogram
for young complexes (i.e. 0.1$<$Q) shows a strong peak of diffuse objects just
inside the spiral (see also Fig.~\ref{n2997xy}), while compact sources are
more evenly distributed.  This is consistent with a density wave scenario in
which a starforming front may form by compression of gas infront of a spiral
potential minimum in the arm \citep{gittins04} for regions inside the
co-rotation radius. Looking at the absolute magnitudes (see
Fig.~\ref{n2997-dth}), only the most massive and brightest complexes with
\MK$<-12.0$~mag are concentrated just inside the arm while less massive
sources are formed also more uniformly in the inter-arm regions.  This
suggests that a compression in the gas associated with a density wave is
necessary to form these very massive complexes.  The abrupt change between
diffuse and compact sources may well be associated with the expulsion of gas
and dust when the first supernovae explode \citep{bastian06, goodwin06}.

The top diagram in Fig.~\ref{n2997-dth} displays the Q index, which for
0.1$<$Q can be considered as an age indicator.  The diffuse sources show a
clear asymmetry where the youngest, around Q=0.4, are furthest away from the
spiral while older ones graduately are closer.  For older complexes with
Q$\approx$-0.1, there are more on the outer side of the spiral
(i.e. $\Delta\theta < 0^\circ$) than inside.  This is in agreement with a
picture where the density wave rotates slower than the material (i.e. inside
co-rotation) and young starforming complexes move out through the arm
from positive to negative values of $\Delta\theta$.  The general asymmetry may
be caused by the short lifetime of young clusters \citep{lada03} but both
statistics and age estimates for older complexes are not sufficient to yield a
reliable estimate.

The $\Delta\theta$-Q gradient for young, diffuse complexes can be used to
estimate the pattern speed $\Omega_\mathrm{p}$ of an underlying density wave
assuming that all complexes are formed in a narrow lane, parallel to the
spiral arm. Using a linear approximation, one obtains:
\begin{equation}
  \Omega_\mathrm{p} = (V_c - V_r/\tan i)/r - \delta\theta/\delta\mathrm{age},
\end{equation}
where $i$ is the pitch angle of the starforming front (negative for a trailing
pattern) while $V_c$ and $V_r$ are the circular and radial velocities,
respectively.  This approximation is valid for $\delta$age $\ll \pi/2\kappa
\approx 50$~Myr for NGC~2997 where $\kappa$ is the epicyclic frequency. The
term $\delta\theta/\delta$age was derived from the $\Delta\theta$-Q gradient
(see top diagram in Fig.~\ref{n2997-dth}) using only the diffuse clusters in
the string along the arm (filled circles) but excluding the 4 clusters (filled
boxes) which lie off the Q-$\Gamma$ relation. This yields an estimate of
0.22$^\mathrm{m}/10^\circ$ (dashed line).  Continuous SB99 models with upper
mass limits above 60~M$_\mathrm{\sun}$ suggest a $\delta$age/$\delta$Q of
-20~Myr per mag for Q values around 0.2~mag (see Fig.~\ref{n2997-cmd}).
Reasonable changes of IMF slope and metallicity for the models do not change
this slope significantly.  The complexes move by 190~pc or 3.9\arcsec\ per Myr
using their average radial distance of $r$ = 3.4~kpc and a circular velocity
of $V_c$ = 185~\kms\ \citep{peterson78}.  This yields an estimate of
$\Omega_\mathrm{p}$ = (54-38) km sec$^{-1}$ kpc$^{-1}$ = 16 km sec$^{-1}$
kpc$^{-1}$ assuming circular motion.  A systematic inward radial velocity of
newly born clusters could be generated by a large-scale shock in the gas and
would lower this estimate.  The circular estimate would place co-rotation at
$r$ = 11.6~kpc = 240\arcsec, significantly outside the strong, symmetric part
of the spiral pattern, and an Inner Lindblad Resonance (ILR) around $r$ =
2.9~kpc = 60\arcsec, close to the start of the main spiral pattern.  These
values of $\Omega_\mathrm{p}$ are slightly lower than previous estimates
\citep{grosbol99, vera01}. The standard SB99 models do not account for Q
values above 0.4 mag as observed and may not predict the NIR colors of the
youngest, most massive clusters well (possibly due to circumstellar
emission). A shallower, upper slope of the IMF would give a slightly better
agreement and increase the $\Omega_\mathrm{p}$ values estimated.  NIR spectra
of the complexes would be required to improve the parameters used for the
models.

\section{Conclusion}
Extended sources in a field centered on the southern arm of NGC~2997 show a
bi-modal distribution with bright, diffuse complexes with \MK $ < -12$~mag
being strongly concentrated along the spiral arm while fainter, more compact
sources are uniformly distributed.  NIR colors suggest that the diffuse
objects in the arm region are starforming complexes with ages $<$8~Myr.  They
display a color gradient as a function of their azimuthal distance from the
spiral arm in the sense that redder, younger complexes are further away from
the arm than bluer, older objects.  It is possible to estimate the pattern
speed of an underlying density wave if one interprets this gradient as an age
variation due the differential motion of material and wave.  Assuming circular
motion, we derive $\Omega_\mathrm{p}$ = 16 km sec$^{-1}$ kpc$^{-1}$ for the
main spiral pattern.  This locates ILR close to the inner limit of the spiral
arms, as seen in K-band images, and co-rotation outside the end of the
symmetric, two-armed pattern.

The strong alignment of young, starforming complexes and their age/color
gradient support a density wave scenario where a large-scale compression of
gas in the arms triggers star formation of the most massive aggregates with
\MK $< -12$~mag.  Young sources fainter than this limit are formed
more uniformly across the disk.

%-----------------------------------------------------------------
\begin{acknowledgements}
  ESO-MIDAS and SExtreator were used for the analysis of the data. HD
  thanks ESO and the Brazilian Council of Research CNPq, Brazil, for support.
  We thank the referee, Dr. I. Perez, for useful comments.
\end{acknowledgements}
%-----------------------------------------------------------------
\bibliographystyle{aa}
\bibliography{AstronRef}

\end{document}